\documentclass[aps,prd,reprint,superscriptaddress]{revtex4-1}

\usepackage[dvips]{graphicx}
\usepackage[utf8]{inputenc}
\usepackage{subfigure}
\usepackage{amssymb}
\usepackage{array,multirow}
\usepackage{amsmath}
\usepackage{mathrsfs}
\usepackage{fancyhdr}
\usepackage{verbatim}
\usepackage{color}

\usepackage{amssymb}
\usepackage{amsthm}
\usepackage{amsmath}

\newcommand{\oo}{\mathcal{O}}

\newcommand{\free}{\mathrm{free}}
\newcommand{\massless}{\mathrm{massless}}

\newcommand{\latt}{\mathrm{latt}}
\newcommand{\MSb}{\overline{\mathrm{MS}}}

\begin{document}

\title{Alleviating the window problem in large volume renormalization schemes}


\author{Piotr Korcyl}
\email[]{piotr.korcyl@ur.de}
\affiliation{Institut f\"ur Theoretische Physik, Universit\"at Regensburg, D-93040 Regensburg, Germany}
\affiliation{M. Smoluchowski Institute of Physics, Jagiellonian University, ul. \L ojasiewicza 11, 30-348 Krak\'ow, Poland}


\date{\today}

\begin{abstract}
We propose a strategy for large volume non-perturbative renormalization which alleviates the window problem by reducing cut-off effects. 
We perform a proof-of-concept study using position space renormalization scheme and the CLS $N_f=2+1$ ensembles generated at 5 different lattice spacings. 
We show that in the advocated strategy results for the renormalization constants are to a large extend independent of the specific lattice direction 
used to define the renormalization condition. Hence, very short lattice distances become accessible even on coarse lattices and the contact with perturbation theory
can be performed at much higher energy scales. Our results include non-perturbatively estimated renormalization constants for quark bilinear operators 
in the scalar, pseudoscalar and axial-vector channels using position space renormalization scheme which we subsequently translate to the $\MSb$ scheme perturbatively 
at $1.5$ GeV. Our proposal is applicable to other non-perturbative large volume renormalization schemes such as RI-MOM and its variants.
\end{abstract}


\maketitle


\section{Introduction}
In the quantum field-theoretical framework of the Standard Model of elementary particles, decay rates of particles are proportional to a kinematic factor multiplied by 
an expectation value of current operator, mediating the particular decay mode, taken between the vacuum and one-particle state. Since 
such hadronic matrix elements are dominated by strong interaction contributions, they can be estimated non-perturbatively, i.e. without resorting to the
perturbative expansion in the coupling constant, using state-of-the-art lattice QCD techniques.
However, bare hadronic matrix elements can not be compared with experimental data. Only renormalized matrix elements have a well defined limit 
when the regulator is removed. In the case of a quantum field theory defined on a space-time lattice that means the continuum limit, i.e. the limit 
of a vanishing lattice spacing $a$. Hence, apart of the hadronic matrix elements, the appropriate renormalization constants need to be calculated separately. 
In this Letter we address the problem of a reliable estimation of renormalization constants using lattice QCD. 

In most cases final 
results are expected to be provided in the $\MSb$ renormalization scheme. As it 
uses dimensional regularization, the $\MSb$ scheme can not be directly combined with lattice results. One has to employ
an intermediate, regularization independent scheme, one example of which is the position space renormalization scheme \cite{Gimenez:2004me,Cichy:2012is} 
described below, and then translate the non-perturbatively determined renormalization constants 
to $\MSb$ at some renormalization scale with a conversion factor usually estimated using continuum massless perturbation theory.

There exists several ways renormalization constants can be determined non-perturbatively on the lattice, see Ref.\cite{Aoki:2010yq} for a review. 
One can either reuse the large volume ensembles generated to estimate hadronic matrix elements or perform separate dedicated simulations
at small volume. In the latter case the most convenient framework is that of Schr\"odinger functional \cite{Luscher:1992an}, as it enables directly simulating massless quarks and hence avoids
performing the chiral extrapolation. The most commonly used large volume renormalization scheme is the RI-MOM scheme \cite{Martinelli:1993dq,Martinelli:1994ty} or its improved versions such as \cite{Sturm:2009kb}, in 
which the renormalization constants are extracted from a specific vertex function at some large momentum. RI-MOM is
regularization independent, allowing for an easy connection to $\MSb$. Alternatively, one can use a position space renormalization scheme \cite{Gimenez:2004me,Cichy:2012is} which uses
correlation functions of particular operators separated by a small distance. Both latter schemes suffer from a window problem, i.e. two inequalities must
be fulfilled on a single lattice: the momentum/distance must be large/small so that the perturbative translation to $\MSb$ can be performed at a scale where perturbation 
theory can be trusted and at the same time the momentum/distance can not be too large/small because of increasing lattice artifacts. 

In this Letter we propose a renormalization strategy which largely suppresses cut-off effects, allowing to use data at very large momenta/short distances even 
at coarse lattice spacings. It works when a set of gauge ensembles covers several lattice spacings and provided at least at one of them the window problem is 
under control. It is applicable for any large volume renormalization scheme, RI-MOM as well as position space. For definiteness 
in the following we concentrate on the position space scheme, because of the ease of implementation.
The rest of the Letter is composed as follows. We start by briefly introducing the position space renormalization scheme. Afterwards, 
we present our proposal and discuss implementation details. We summarize our results in tables \ref{tab. za}-\ref{tab. zsb} by providing estimates 
of the renormalization constants in the scalar, pseudoscalar and axial channels, which are relevant in the determination of renormalized quark masses. 
We compare our results with those obtained using Schr\"odinger Functional scheme for $Z_A$. Estimates 
of $Z^{\MSb}_S(2 \ \textrm{GeV})$ and $Z^{\MSb}_P(2 \ \textrm{GeV})$ are new for the gauge ensembles considered here. 
Last section contains some conclusions.

\section{Position space renormalization scheme}
\label{sec. xspace}

Position space renormalization scheme has the advantage of being constructed with gauge-invariant and on-shell correlation functions, hence avoids 
completely many problems such as the gauge fixing procedure, possible contamination with Gribov copies or mixing with additional operators which would otherwise vanish 
by evoking the equations of motion. In the following we consider correlation function of flavor non-singlet bilinear quark operators of the form
\begin{equation}
C_{\Gamma}(x) = \langle \bar{\psi}^i(x) \Gamma \psi^j(x) \bar{\psi}^i(0) \Gamma \psi^j(0) \rangle, 
\end{equation}
with $\Gamma = \{ 1, \gamma_5, \gamma_{\mu} \gamma_5 \}$ and $i \ne j$. The renormalization condition is imposed at a given physical distance $x_0$ by
equating the non-perturbatively estimated correlation function to its tree-level value,
\begin{equation}
 \lim_{a\rightarrow 0}\langle \oo^X_{\Gamma}(x) \oo^X_{\Gamma} (0) \rangle
\big|_{x^2=x_0^2} = \langle \oo_{\Gamma}(x_0) \oo_{\Gamma}(0)
\rangle^{\free, \massless}_{\latt},
\label{eq. condition}
\end{equation}
where on the right hand side we used the tree-level value estimated in the lattice perturbation theory to cancel the leading tree-level cut-off effects 
as was advocated in Ref.\cite{Gimenez:2004me,Cichy:2012is}. Then, the renormalized operator in the X-scheme is 
\begin{equation}
 \oo^X_{\Gamma}(x, x_0) = Z^X_{\Gamma}(x_0) \oo_{\Gamma}(x), 
\end{equation}
with the renormalization constant given by
\begin{equation}
Z_{\Gamma}^X(x_0) = \sqrt{\frac{C(x_0)^{\free, \massless}_{\Gamma,\latt}}{C_{\Gamma}(x_0)}}.
\label{eq. z}
\end{equation}
The resulting renormalization constants in the position space scheme can be translated to the $\MSb$ scheme with a perturbative factor; for the case of
quark bilinear operators this translation factor is known up to $\alpha_S^4$ \cite{Chetyrkin:2010dx}.

As any large volume renormalization scheme, this scheme suffers from the window problem, namely the specific lattice distance
$x_0$ must satisfy the inequality
\begin{equation}
 a\ll x_0 \ll \Lambda^{-1}_{\mathrm{QCD}} 
\end{equation}
in order to keep the discretization $( a \ll x_0)$ and non-perturbative effects $( x_0 \ll \Lambda^{-1}_{\mathrm{QCD}})$ under control. Such
a window will ultimately exist for lattices with a lattice spacing fine enough and it is an empirical question which needs to be addressed in each 
particular problem separately, whether such a window exists for lattice spacings employed in the given simulation. The proposal which we
describe in the following, weakens the lower requirement, $a \ll x_0$, by allowing lattice distances $x_0 \approx a$ without introducing large
lattice artifacts.

\section{New renormalization strategy}
\label{sec. strategy}

\begin{figure}
\begin{center}
\includegraphics[width=0.45\textwidth]{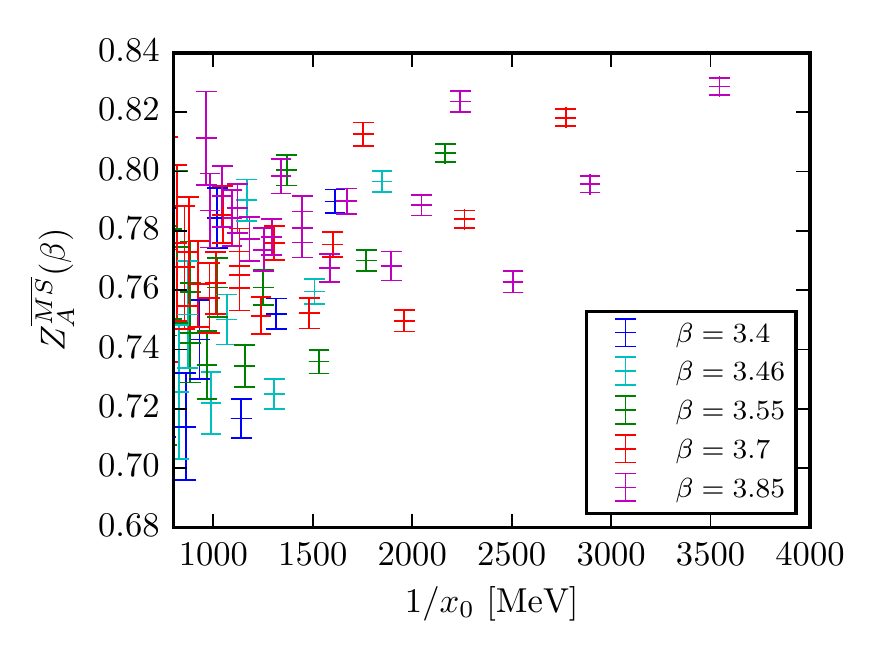}
\caption{Estimates of $Z_A$ obtained using Eq.\ref{eq. z} and translated to the $\MSb$ scheme. Large lattice artifacts are responsible for the scatter of data points. One can improve 
on that by keeping only the so-called democratic points which minimize the tree-level artifacts. Note that the pattern of the scatter is very similar for all
values of $\beta$. \label{fig. traditional}}
\end{center}
\end{figure}

\begin{figure}
\begin{center}
\includegraphics[width=0.45\textwidth]{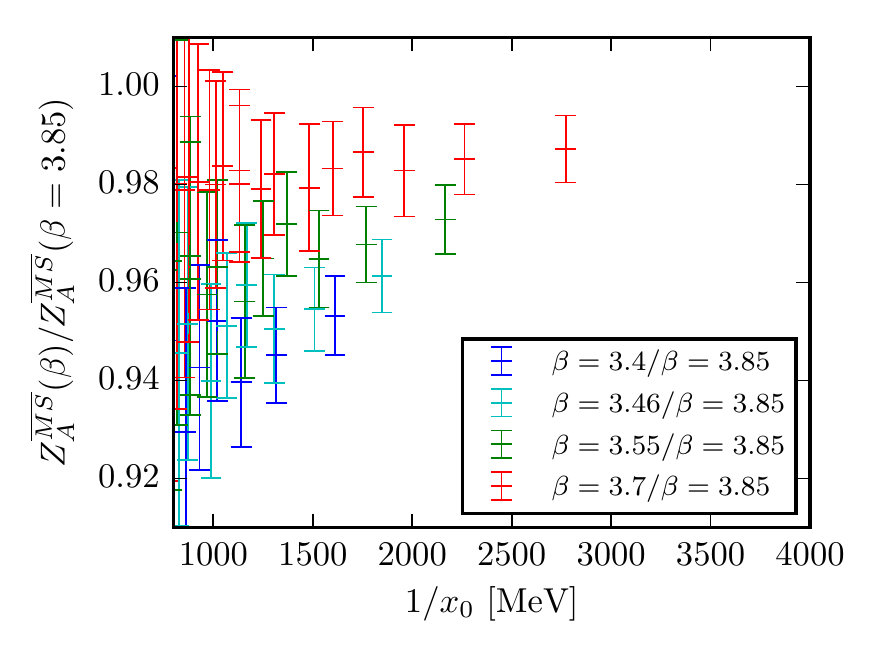}
\caption{Estimates of $Z_A(\beta)/Z_A(\beta=3.85)$. Large lattice artifacts cancel in the ratio giving a clear signal for the value of the ratio, 
even at very short distances. Data are shifted by a constant offset for better readability. \label{fig. ratios}}
\end{center}
\end{figure}

Our proposal relies on two observations. Firstly, the renormalization constant extracted at the finest available lattice
spacing is the least affected by cut-off effects. In other words, the window problem for the ensemble with the finest lattice spacing 
is the least severe as the inequalities are most likely to be fulfilled at some lattice distance $n_0$, $1 \ll n_0 \ll 1/(a\Lambda)$. 
The second observation states that cut-off effects mostly cancel in a ratio of renormalization constants determined at the same lattice distance, $n$. 
On figure \ref{fig. traditional} where the naive data is plotted for five available lattice spacings, one notices that the pattern of the scatter
of data points is similar among the different values of $\beta$. When the ratio of renormalization constants for different lattice spacings but at the same lattice distance
is plotted, the scatter disappears, see figure \ref{fig. ratios}. Hence, employing both observations allows to rewrite the renormalization constant at any $\beta$ using 
a factorization in which in each term cut-off effects are under control. 

In order to make it more concrete, assume that we dispose of two
ensembles: one at an inverse coupling constant $\beta$ with a coarse lattice spacing and the second with a fine lattice spacing given by $\hat{\beta}$. 
Then, the renormalization constant at a given value of the inverse coupling 
constant $\beta$ and renormalization scale $\mu$, $Z_J(\beta, a\mu)$ can be factorized into a renormalization constant evaluated at the finest available
lattice spacing, given by $\hat{\beta}$ and at a renormalization scale $\hat{\mu}$, $Z_J(\hat{\beta}, \hat{a}\hat{\mu})$ and a ratio which describes the running
of $Z_J$ from $\hat{\beta}$ to $\beta$ which can be estimated non-perturbatively and in which most of the cut-off effects cancel,
\begin{equation}
\label{eq.}
Z_J(\beta, a\mu=1/n) = \hat{Z}_J(\hat{\beta}, \mu') \frac{Z_J(\beta, 1/n)}{Z_J(\hat{\beta}, 1/n)},
\end{equation}
where
\begin{equation}
\hat{Z}_J(\hat{\beta}, \mu') = Z_J(\hat{\beta}, \hat{a}\hat{\mu}=1/n_0) R(\hat{a}\mu'=1/n, \hat{a}\hat{\mu}=1/n_0),
\end{equation}
and $R(\mu', \hat{\mu})$ is a perturbative factor describing the running of $Z_J$ from the scale $\hat{\mu}$ to $\mu' = 1/(\hat{a} n)$. 
The factor $R$ is expected to be well described by perturbation theory because the involved scales
$\mu'$ and $\hat{\mu}$ are relatively large. The last term is a ratio of renormalization constants taken at different
values of $\beta$ but at the same lattice distance $n$. Lattice artifacts cancel to a large extend in this ratio, and therefore its value describes the non-perturbative
change of the renormalization constant when the lattice spacing is varied. The combination $Z_J(\hat{\beta}, \hat{a}\hat{\mu}=1/n_0) R(\mu', \hat{\mu})$ is
chosen in such a way as to cancel the denominator of the ratio. What remains is the renormalization constant at the coarse lattice spacing $\beta$ with largely reduced 
lattice artifacts. As is demonstrated below, this idea allows to use lattice distances of type (0,0,1,1) at the lattice spacing of $0.086$fm. As a consequence, 
in the investigated setup the renormalization condition can be applied at a small but fixed physical distance, say $1.5$ GeV, for all lattice spacings without
introducing uncontrollable cut-off effects.

\section{Application}
\label{sec. application}

\subsection{Gauge ensembles and evaluation of correlation functions}

The ensembles used in this study are summarized in table \ref{tab. ensembles}.
They cover a range of five lattice spacings from 0.086 fm down to 0.039 fm and with pion masses ranging from 712 Mev to 229 MeV. On each ensemble we measured
the following correlation functions
\begin{align}
G_{S^{(jk)}}(x) &= \left\langle S^{(jk)}(x)
\overline{S}^{(jk)}(0) \right\rangle
 \label{eq corr s 12} \\
G_{P^{(jk)}}(x) &=
\left\langle P^{(jk)}(x)
\overline{P}^{(jk)}(0) \right\rangle
\label{eq corr p 12} \\
\label{eq corr a 12}
G_{A^{(jk)}}(x) &= \frac{1}{4} \sum_{\mu}
\left\langle A_{\mu}^{I, (jk)}(x)
\overline{A}_{\mu}^{I, (jk)}(0) \right\rangle\\\nonumber
\end{align}
We non-perturbatively improved the axial current using $c_A$ from Ref.\cite{Bulava:2015bxa}.
In order to decrease the statistical errors we employed the Truncated Solver Method \cite{Bali:2009hu}. We used the IDLFS solver from the Ref.\cite{Luscher:2007se} 
implemented in Chroma \cite{Edwards:2004sx}. 
On each configuration we performed 64 solves with the maximal iteration count reduced to 6 iterations and 4 exact solves in order to estimate the systematic bias of the 
truncated solves. We found that at short distances, $|x|\le6$, the bias is at the level of $10^{-9}$ and is completely negligible when compared with the statistical noise.
For each lattice distance $x$ we averaged the correlation functions over all sites equivalent over the $H_4$ symmetry.

\subsection{Chiral extrapolation}

\begin{table}
\begin{center}
\begin{ruledtabular} 
\begin{tabular}{cccccc}
$\beta$ & name & $\kappa_l = \kappa_s$ & $m_{\pi}$ [MeV] & $t_0/a^2$ & \# conf. \\
\hline
3.4  & H101   & 0.136759 & 416 & 2.8468(59) & 372 \\
3.4  & rqcd17 & 0.136865 & 229 & 3.2509(93) & 457 \\
3.4  & rqcd21 & 0.136813 & 336 & 3.0318(15) & 387 \\
3.46 & B450   & 0.136890 & 421 & 3.671(18)  & 320 \\
3.46 & H400   & 0.136888 & 420 & 3.630(21)  & 189 \\
3.46 & rqcd29 & 0.136600 & 712 & 2.9758(47) & 434 \\
3.46 & rqcd30 & 0.136959 & 323 & 3.9152(77) & 280 \\
3.55 & B250   & 0.136700 & 706 & 5.873(8)   & 84  \\
3.55 & N202   & 0.137000 & 410 & 5.164(16)  & 177 \\
3.55 & X250   & 0.137050 & 347 & 5.2818(13) & 182 \\
3.55 & X251   & 0.137100 & 268 & 5.4831(14) & 177 \\
3.7  & N300   & 0.137000 & 416 & 8.576(30) & 197 \\
3.7  & N303   & 0.136800 & 630 & 7.739(32) & 99  \\
3.85 & J500   & 0.136852 & 396 & 13.972(32) & 120 \\
\end{tabular}
\end{ruledtabular} 
\end{center}
\caption{Summary of $N_f=2+1$ CLS ensembles at the symmetric line $\kappa_l=\kappa_s$ used in this work. The gauge action is the tree-level Symanzik
improved action, while the fermion action is discretized with the Wilson O(a)-improved action with $c_{SW}$ determined non-perturbatively. All ensembles
except rqcd?? and X25? feature open boundary conditions in the time direction, the latter have antiperiodic boundary conditions. For more details see 
Refs.\cite{Bruno:2014jqa,Bali:2016umi}. The values of $t_0/a^2$ are the reweighted estimates using the symmetric definition of the Yang-Mills action density. 
The ensemble with the finest lattice spacing is the J500 ensemble at $\beta=3.85$ with a lattice spacing of $0.0391(15)$ fm. 
Pion masses were taken from Ref.\cite{Bali:2016umi} and from Wolfgang S\"oldner (private communications). The last column indicates the size of the subset 
of available configurations used for measurements. \label{tab. ensembles}}
\end{table}

\begin{table}
\begin{center}
\begin{ruledtabular} 
\begin{tabular}{cccccc}
$\beta$ & name & $\kappa_{\textrm{see}}$ &  $\kappa_{\textrm{valence}}$ & $m_{\pi}$ [MeV] & $t_0/a^2$ \\
\hline
3.55 & \emph{X250a} & 0.137050 & 0.137100 & 300 & 5.2818(13) \\
3.7  & \emph{N300a} & 0.137000 & 0.137050 & 330 & 8.576(30) \\
3.85 & \emph{J500a} & 0.136852 & 0.136700 & 340 & 13.972(32) \\
3.85 & \emph{J500b} & 0.136852 & 0.136750 & 250 & 13.972(32) \\
\end{tabular}
\end{ruledtabular} 
\end{center}
\caption{Summary of non-unitary measurements: valence quark masses have different values that the sea quarks masses. The name refers to the 
name of the ensemble from which the configurations were taken. \label{tab. nonunitary}}
\end{table}

\begin{figure*}
\begin{center}
\includegraphics[width=0.45\textwidth]{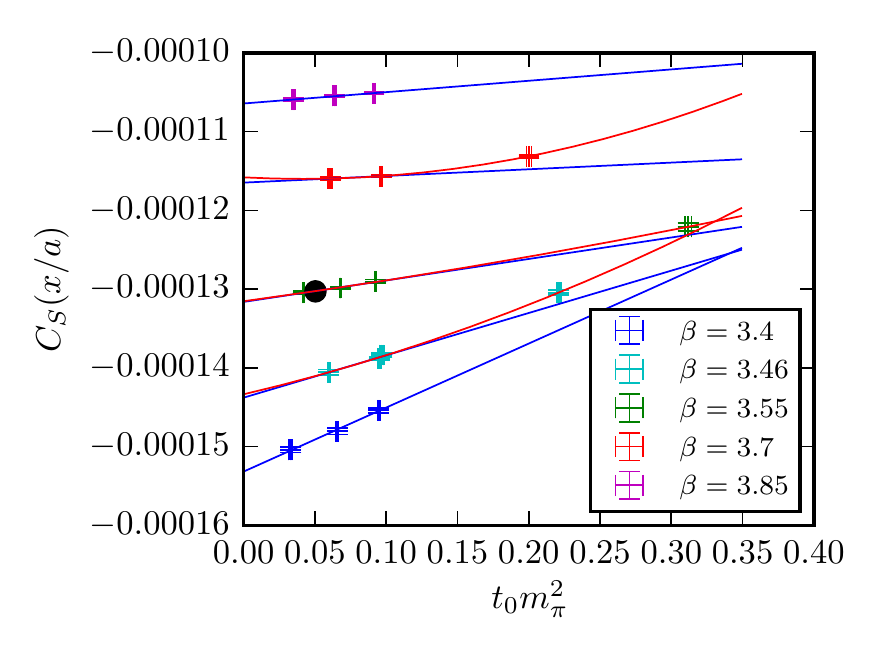}
\includegraphics[width=0.45\textwidth]{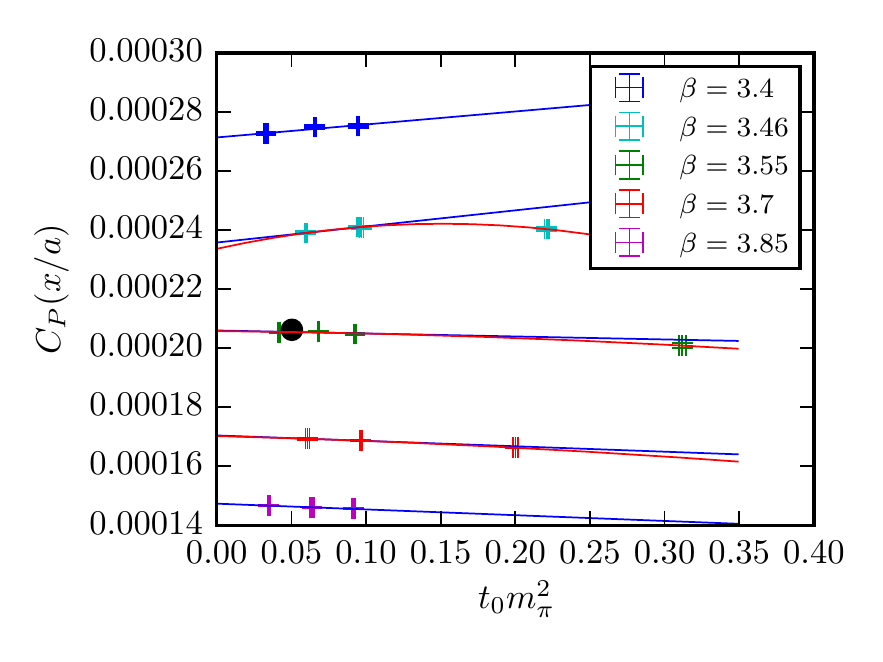}
\caption{Examples of chiral extrapolation of $C_S(x)$ and $C_P(x)$ at an intermediate $x=(0,1,2,2)$. 
Statistical uncertainties on particular data points may be smaller than the symbol. The black data point was not included in the fit and serves as an estimate of
the systematic uncertainty of the non-unitary measurements.\label{fig. extrapolations}}
\end{center}
\end{figure*}

The renormalization condition Eq. \eqref{eq. condition} is imposed in the chiral limit, as we are interested in a massless renormalization scheme. 
Hence, the correlator data measured along the symmetric line $\kappa_l = \kappa_s$ have to be extrapolated to the point where the pion mass vanishes. We
parametrized our data with a dimensionless variable $y = t_0 m_{\pi}^2$ proportional to the renormalized quark mass. Most of our data 
has $y < 0.1$ and shows a linear dependence on $y$. For $\beta=3.46, 3.55$ and $\beta=3.7$ we have additional measurements for heavier pion masses. They no
longer show a linear dependence on $y$, rather they can be described by an additional term quadratic in $y$. The chiral extrapolation is performed independently for each
value of $\beta$ with a linear Ansatz for data satisfying $y < 0.1$ and a linear plus quadratic Ansatz for cases where additional data exist. Their difference is taken
as an estimate of the systematic uncertainty of the extrapolation. The fits include uncertainties along both vertical and horizontal axes. We assume them to be statistically 
uncorrelated, as the pion mass is extracted from the slope of the correlator and its fluctuations can be considered independent from the fluctuations of the overall offset.

At the current stage of gauge ensemble generation only a single ensemble exists on the symmetric line with $y< 0.1$ for the two finest lattice spacings, namely $\beta=3.7$ and $\beta=3.85$. 
For the purpose of this proof-of-concept study we supplemented our data with three non-unitary measurements, marked in italics in table \ref{tab. nonunitary},
which have different $\kappa$ values for the valence and sea quarks. This approximation will not be necessary once more ensembles are available 
for these values of $\beta$ in the near future. For each of these additional measurements we estimated the correlators as well as the new pion masses; the value
of $t_0/a^2$ being unchanged. In order to estimate the systematic effect introduced by the non-unitary setup we performed another non-unitary measurement at $\beta=3.55$. This
particular data point is not included in the fits. Rather, its deviation from the predicted fit was estimated and turned out to be statistically 
insignificant. Two examples of chiral fits are shown on figure \ref{fig. extrapolations}, with the additional data point at $\beta=3.55$ marked in black.

\subsection{Error analysis}

The total error budget is composed of the following parts: statistical uncertainty, uncertainty of $\Lambda_{\textrm{QCD},\MSb}^{(3)}$, uncertainty of the scale setting and
the systematic uncertainty of the chiral extrapolation. As far as the the statistical errors are concerned, we propagated the uncertainties of all individual data points 
to the uncertainties of the fit parameters through a bootstrap procedure with an ensemble of 2000 samples. 
Statistical correlations between the starting data point at $\beta=3.85$ and the ratio capturing the running to other values of $\beta$ is taken into account 
in the bootstrap procedure. Autocorrelations turned out to be negligible for the correlation
functions at question, as long as only the short distance part was used. The measured $\tau_{\textrm{int}}$ was below 1.0. 
The systematic uncertainty of the chiral extrapolation is quantified by the
difference between the results obtained with the linear and quadratic fit functions. 
In order to compute the running factor 
$R(\mu, \hat{\mu})$ we used perturbative expressions estimated up to and including terms $\alpha_s^4$ from Ref.\cite{Chetyrkin:2010dx}.
We used the value of $\Lambda_{\textrm{QCD},\MSb}^{(3)} = 336(19)$ taken from Ref.\cite{Aoki:2016frl}. The lattice spacings in physical 
units were taken from Ref.\cite{Bruno:2016plf}. In all cases the final uncertainties were dominated by 
systematic uncertainties originating in the uncertainties of the lattice spacing and of the $\Lambda^{(3)}_{\textrm{QCD},\MSb}$ parameter. 
The statistical errors as well as the chiral extrapolation uncertainty were at least one order of magnitude smaller. 


\section{Results}
\label{sec. results}

\begin{figure}
\begin{center}
\includegraphics[width=0.45\textwidth]{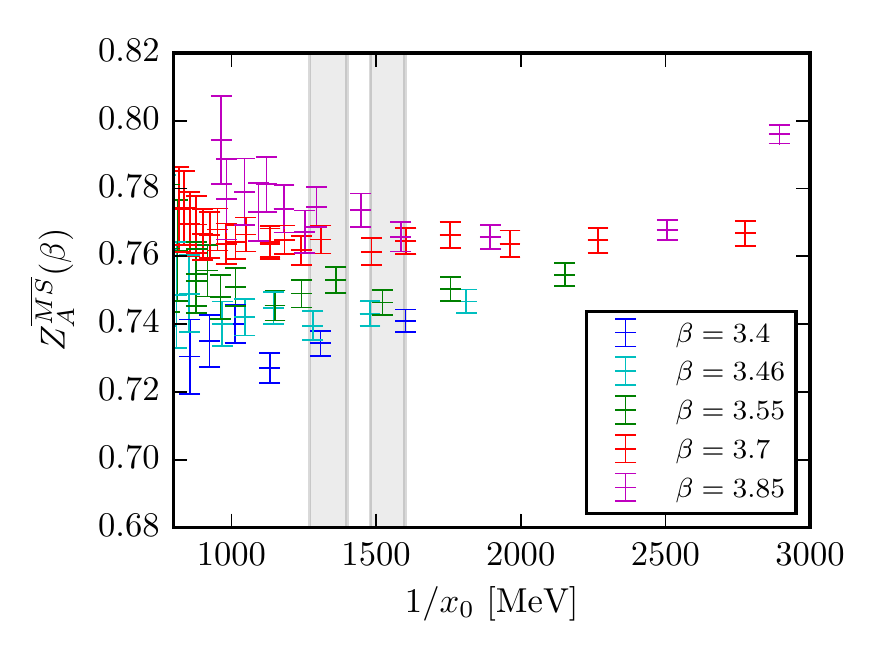}
\caption{Results obtained with the proposed strategy. The grey area encompassing the scale interval $[1.27,1.40]$ GeV $\approx 1.3$ GeV contains data points
used in strategy A, whereas the grey area encompassing the scale interval $[1.48,1.60]$ GeV $\approx 1.5$ GeV contains data points
used in strategy B. \label{fig. results}}
\end{center}
\end{figure}

Our results include estimates of the renormalization constants, $Z^{\MSb}_P(2 \ \textrm{GeV})$, $Z^{\MSb}_S(2 \ \textrm{GeV})$ and $Z_A^{\MSb}$ for 5 values of $\beta$. 
On figure \ref{fig. results} we demonstrate the results for $Z_A$ obtained with the proposed strategy. We start with the determination of the renormalization constant at the 
finest lattice spacing, i.e. at $\beta=3.85$ using strategy A (see table \ref{tab. distances}). We choose $n_0$ to be $(2,2,2,2)$ as it corresponds to an energy scale 
of 1.253 GeV and $|n_0| \gg 1$. Had we chosen a neighboring data point, the final results would not change, since in this region all data points 
are compatible within their errors. In other words, for this lattice spacing and this renormalization constant, the window problem is mild. Subsequently, 
at each of the four remaining lattice spacings we evaluated the renormalization constants following Eq.\ref{eq.} using different lattice directions $n$ and 
included the necessary perturbative factor $R$. One notices that the scatter of data points at these values of $\beta$ is much reduced. At the coarsest available 
lattice spacing at $\beta=3.4$ one can extract the value of the renormalization constant using lattice distance of $n=(0,0,1,1)$, which was clearly not possible before. 
Final results were calculated by imposing the renormalization condition at a single, physical scale of 1.3 GeV for all five lattice spacings by taking the lattice 
distances listed in table \ref{tab. distances}. For demonstration purposes, we also defined a second set of lattice distances corresponding to approximately 1.5 GeV, 
which we called choice B. In this strategy we start with a lattice distance $n_0=(1,1,2,2)$ at $\beta=3.85$ and proceed along Eq.\ref{eq.} at other values of $\beta$. 
The differences between renormalization constants obtained using both choices give an estimate of remaining cut-off effects. We discuss the particular channels in 
the next subsections.

\begin{table}
\begin{center}
\begin{ruledtabular} 
\begin{tabular}{ccccc}
& \multicolumn{2}{c}{choice A}& \multicolumn{2}{c}{choice B} \\
\hline
$\beta$ & distance $n$ & scale [GeV]  & distance $n$ & scale [GeV] \\
\hline
3.4  & (0,1,1,1) & 1.308 & (0,0,1,1) & 1.601 \\
3.46 & (1,1,1,1) & 1.281 & (0,1,1,1) & 1.479 \\
3.55 & (0,0,1,2) & 1.361 & (1,1,1,1) & 1.522 \\
3.7  & (0,1,2,2) & 1.309 & (0,1,1,2) & 1.603 \\
3.85 & (2,2,2,2) & 1.253 & (1,1,2,2) & 1.585 \\
\end{tabular}
\end{ruledtabular}
\end{center}
\caption{Summary of lattice distances used to impose the renormalization condition.\label{tab. distances}}
\end{table}

\begin{figure}
\begin{center}
\includegraphics[width=0.45\textwidth]{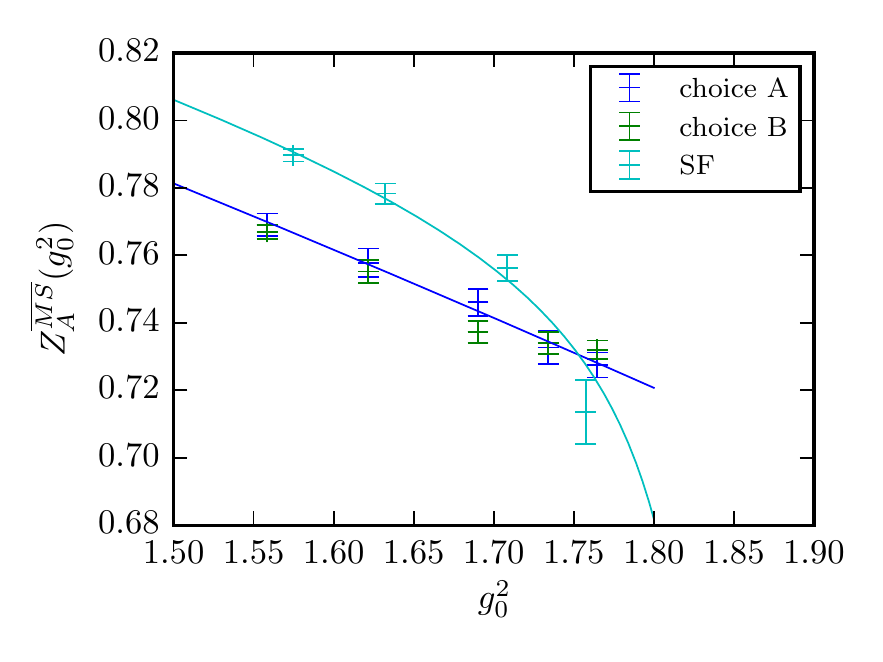}
\caption{Comparison with the results for $Z_A$ obtained using the Schr\"odinger Functional renormalization scheme \cite{Bulava:2016ktf}. \label{fig. sf}}
\end{center}
\end{figure}

\subsection{Axial current renormalization constant}

Final results for $Z_A^{\MSb}$ are listed in table \ref{tab. za}. On figure \ref{fig. sf} we compare our outcomes using both choice A and B of the 
renormalization conditions as well as with Ref.\cite{Bulava:2016ktf} which used Schr\"odinger Functional renormalization scheme. The 
three renormalization schemes differ by $\mathcal{O}(a^2)$ cut-off effects and we find that values lie relatively close to the 
parametrization provided in that Reference. Parametrizing our data with 
a functional Ansatz with two fit parameters, $a$ and $b$, and a constrained one-loop coefficient \cite{Aoki:1998ar},
\begin{equation}
f(g_0^2) = 1-0.090488 g_0^2 \ \frac{1-a \ g_0^2}{1-b \ g_0^2}
\end{equation}
we find the values of $a=-0.570$, $b=-0.101$ and of $0.47$ for the $\chi^2$ per degree of freedom. The resulting parametrization is plotted on figure \ref{fig. sf}.

\begin{table}
\begin{center}
\begin{ruledtabular} 
\begin{tabular}{ccccccc}
$\beta$ & $Z^{\MSb}_A$ & stat. & $\Lambda_{\textrm{QCD}}^{(3)}$  & $a$ & chiral & total \\
\hline
3.4  & 0.72753 & 0.00192 & 0.00309 & 0.00123 & --- & 0.00384  \\
3.46 & 0.73267 & 0.00294 & 0.00313 & 0.00121 & 0.00224 & 0.00499  \\
3.55 & 0.74608 & 0.00262 & 0.00286 & 0.00111 & 0.00073 & 0.00409  \\
3.7  & 0.75783 & 0.00316 & 0.00271 & 0.00091 & 0.00078 & 0.00433  \\
3.85 & 0.76908 & 0.00149 & 0.00207 & 0.00140 & --- &  0.00291 \\
\hline
3.4  & 0.73205 & 0.00189 & 0.00186 & 0.00068 & --- & 0.00273  \\
3.46 & 0.73410 & 0.00231 & 0.00196 & 0.00065 & 0.00107 & 0.00328  \\
3.55 & 0.73734 & 0.00253 & 0.00182 & 0.00058 & 0.00059 & 0.00322  \\
3.7  & 0.75521 & 0.00298 & 0.00159 & 0.00046 & 0.00063 & 0.00346  \\
3.85 & 0.76691 & 0.00137 & 0.00127 & 0.00086 & --- &  0.00206 \\
\end{tabular}
\end{ruledtabular}
\end{center}
\caption{Summary of results for the axial renormalization constant using strategy A (upper part) and strategy B (lower part).\label{tab. za}}
\end{table}

\subsection{Scalar and pseudoscalar currents renormalization constants}

\begin{table}
\begin{center}
\begin{ruledtabular} 
\begin{tabular}{ccccccc}
$\beta$ & $Z^{\MSb}_S(2 \ \textrm{GeV})$ & stat. & $\Lambda_{\textrm{QCD}}^{(3)}$  & $a$ & chiral & total \\
\hline
3.4  & 0.5716 & 0.0019 & 0.0081 & 0.0078 & ---    & 0.0169 \\
3.46 & 0.5580 & 0.0032 & 0.0075 & 0.0071 & 0.0015 & 0.0161 \\
3.55 & 0.5692 & 0.0023 & 0.0053 & 0.0063 & 0.0000 & 0.0138 \\
3.7  & 0.5564 & 0.0032 & 0.0032 & 0.0041 & 0.0005 & 0.0111 \\
3.85 & 0.5395 & 0.0013 & 0.0011 & 0.0113 & ---    & 0.0131 \\
\hline
3.4  & 0.5823 & 0.0018 & 0.0053 & 0.0079 & ---    & 0.0097 \\
3.46 & 0.5639 & 0.0027 & 0.0053 & 0.0072 & 0.0010 & 0.0094 \\
3.55 & 0.5471 & 0.0023 & 0.0038 & 0.0063 & 0.0002 & 0.0077 \\
3.7  & 0.5500 & 0.0028 & 0.0015 & 0.0049 & 0.0008 & 0.0059 \\
3.85 & 0.5381 & 0.0012 & 0.0012 & 0.0113 & ---    & 0.0114 \\
\end{tabular}
\end{ruledtabular} 
\end{center}
\caption{Summary of results for $Z_S^{\MSb}(2 \textrm{GeV})$ using strategy A (upper part) and strategy B (lower part).\label{tab. zsa}}
\end{table}

\begin{table}
\begin{center}
\begin{ruledtabular} 
\begin{tabular}{ccccccc}
$\beta$ & $Z^{\MSb}_P(2 \ \textrm{GeV})$ & stat. & $\Lambda_{\textrm{QCD}}^{(3)}$  & $a$ & chiral & total \\
\hline
3.4  & 0.4605 & 0.0014 & 0.0065 & 0.0063 & ---    & 0.0141  \\
3.46 & 0.4574 & 0.0022 & 0.0062 & 0.0059 & 0.0014 & 0.0136  \\
3.55 & 0.4715 & 0.0018 & 0.0044 & 0.0052 & 0.0002 & 0.0120  \\
3.7  & 0.4653 & 0.0025 & 0.0027 & 0.0035 & 0.0019 & 0.0096  \\
3.85 & 0.4648 & 0.0010 & 0.0010 & 0.0097 & ---    & 0.0114  \\
\hline
3.4  & 0.4741 & 0.0013 & 0.0044 & 0.0065 & ---    & 0.0080  \\
3.46 & 0.4650 & 0.0020 & 0.0044 & 0.0059 & 0.0013 & 0.0077  \\
3.55 & 0.4561 & 0.0018 & 0.0032 & 0.0052 & 0.0001 & 0.0064  \\
3.7  & 0.4655 & 0.0022 & 0.0013 & 0.0041 & 0.0006 & 0.0049  \\
3.85 & 0.4651 & 0.0010 & 0.0010 & 0.0097 & ---    & 0.0098  \\
\end{tabular}
\end{ruledtabular} 
\end{center}
\caption{Summary of results for $Z_P^{\MSb}(2 \textrm{GeV})$ using strategy A (upper part) and strategy B (lower part).\label{tab. zsb}}
\end{table}

We collect the values for the scalar and pseudoscalar renormalization constants 
for the $N_f=2+1$ CLS ensembles in the $\MSb$ scheme at 2 GeV in tables \ref{tab. zsa} and \ref{tab. zsb}, for both strategies. 
Although these strategies use different lattice distances $n$ to define the renormalization conditions, the outputs are compatible within their errors.

Additionally we tabulate data for the ratio $Z_S/Z_P$ and the combination $Z = Z_P/(Z_S Z_A)$. Both of them are renormalization scale independent 
and therefore free of the uncertainty coming from $\Lambda_{\textrm{QCD},\MSb}^{(3)}$. In tables \ref{tab. zszp} and \ref{tab. z} 
we present results obtained using both strategies, as a demonstration of remaining cut-off effects. We note that in Ref.\cite{Bali:2016umi} the combination $Z$ 
was determined through a global fit for $\beta=3.4$ and $\beta=3.55$ to PCAC-like relations giving a slightly smaller values. This discrepancy
can be probably attributed to the remaining $\mathcal{O}(am)$ cut-off effects in that Reference.

\begin{table}
\begin{center}
\begin{ruledtabular} 
\begin{tabular}{ccc}
$\beta$ & choice A & choice B \\
\hline
3.4 & 1.241(08) & 1.229(07)  \\
3.46 & 1.220(16) & 1.213(13) \\
3.55 & 1.207(10) & 1.199(10) \\
3.7 & 1.196(15) & 1.182(12)  \\
3.85 & 1.161(06) & 1.157(05) \\
\end{tabular}
\end{ruledtabular} 
\end{center}
\caption{Summary of results for the ratio $Z_S/Z_P$.\label{tab. zszp}}
\end{table}

\begin{table}
\begin{center}
\begin{ruledtabular} 
\begin{tabular}{cccc}
$\beta$ & choice A & choice B & Ref.\cite{Bali:2016umi} \\
\hline
3.4 & 1.107(13) & 1.111(11) & 0.8710(32) \\
3.46 & 1.119(22) & 1.123(17) & --- \\
3.55 & 1.110(15) & 1.131(14) & 0.9841(25) \\
3.7 & 1.103(21) & 1.121(17) & --- \\
3.85 & 1.120(10) & 1.127(08) & --- \\
\end{tabular}
\end{ruledtabular} 
\end{center}
\caption{Summary of results for the combination $Z=Z_P/(Z_S Z_A)$.\label{tab. z}}
\end{table}

\section{Conclusions}
\label{sec. conclusions}

In this Letter we proposed a strategy for non-per\-tur\-ba\-ti\-ve renormalization in large volume which suppresses the contamination of 
extracted renormalization constants by lattice artifacts. It is applicable when data for several lattice spacings are available as 
is usually the case when a continuum extrapolation of lattice data is attempted. We applied it to the CLS ensembles featuring 5 lattice spacings and
evaluated the scalar, pseudoscalar and axial renormalization constants. This set of renormalization constants is particularly interesting because 
it can be used to estimate renormalized quark masses through PCAC-like relations. We obtained a total uncertainty of 0.5\% for $Z_A$ and about 3\% for $Z_S$ and $Z_P$. 
We are currently investigating a further improvement which is the estimation and subtraction of one-loop lattice artifacts.  

\begin{acknowledgments}
The Author thanks Agnieszka Kujawa-Cichy for providing a C implementation of the formulae of the renormalization constants running from 
Ref.\cite{Chetyrkin:2010dx}, Gunnar Bali for encouragement, discussions and reading the manuscript, Krzysztof Cichy for many 
helpful discussions and Wolfgang S\"oldner for measuring several pion masses 
as well as all colleagues from the CLS initiative. 
This research was carried out with the support of the Interdisciplinary Centre for Mathematical and Computational
Modelling (ICM) University of Warsaw under grant No. GA67-12. This work was supported by Deutsche Forschungsgemeinschaft under Grant No. SFB/TRR 55
and in part by the polish NCN grant No. UMO-2016/21/B/ ST2/01492.

The gauge ensembles were generated with the help of the Gauss Centre for
Supercomputing e.V. (http://www.gauss-centre.eu) using computer time allocations 
on SuperMUC at Leibniz Supercomputing Centre (LRZ, http://www.lrz.de) and
JUQUEEN at Jülich Supercomputing Center (JSC, http://
www.fz-juelich.de/ias/jsc).
GCS is the alliance of the three national supercomputing centers HLRS (Universität
Stuttgart), JSC (Forschungszentrum Jülich) and LRZ
(Bayerische Akademie der Wissenschaften), funded by
the German Federal Ministry of Education and Research
(BMBF) and the German State Ministries for Research of
Baden-Württemberg (MWK), Bayern (StMWFK) and
Nordrhein-Westfalen (MIWF). 
Additionally computer time provided by PRACE
(Partnership for Advanced Computing in Europe, http://
www.prace-ri.eu) as part of the project ContQCD was used.
Additional simulations were performed on the
Regensburg iDataCool cluster and on the SFB/TRR 55
QPACE computer \cite{Baier:2009yq}, \cite{NAKAMURA2011841}. OPENQCD \cite{Luscher2013519} was used to
generate the main gauge ensembles, as part of the joint CLS
effort \cite{Bruno:2014jqa}. Additional ensembles were generated
on QPACE (using BQCD \cite{NAKAMURA2011841},\cite{Hoelbling:2011kk}) and on the Wilson HPC
Cluster at IKP Mainz.
\end{acknowledgments}
\bibliography{references2}

\end{document}